# Selective Atomic Hydrogen Heating in Plasmas: Implications for Quantum Theory

J Phillips

University of New México, Department of Mechanical Engineering,
Albuquerque, NM 87103

**Abstract.** A new model of quantum mechanics, Classical Quantum Mechanics, is based on the (nearly heretical) postulate that electrons are physical objects that obey classical physical laws. Indeed, ionization energies, excitation energies etc. are computed based on picturing electrons as 'bubbles' of charge that symmetrically surround a nucleus. Hence, for example, simple algebraic expressions based on Newtonian force balances are used to predict ionization energies and stable excitation states with remarkable precision. One of the most startling predictions of the model is that there are stable 'sizes' of the hydrogen atom electron (bubble diameter) that are smaller ('hydrinos') than that calculated for the 'ground state'. Experimental evidence in support of this novel physical/classical version of quantum is alleged to be found in the existence of super heated hydrogen atoms reported by many teams in a variety of plasmas. It is postulated that the energy required for creating super heated H atoms comes from the shrinkage of ground state H atoms to form hydrinos. This claim is discussed with reference to a brief review of the published studies of selective Balmer series line broadening in pure $H_2$ and mixed gas plasmas.

## 1. Introduction

Paradigms once firmly established are not only difficult to replace, they are difficult to question. For example, the current paradigm of quantum mechanics, that electrons, both bound and free, can be 'described' by a 'wave function' derived from a solution to Shrodinger's equation, is so deeply embedded in physics education and culture that 'belief' in it is ultimate truth is virtually a requirement of someone who hopes to be accepted as a member in good standing of the physics community. Don't even try to get a research grant for exploring alternative paradigms!

Yet there is a serious challenge to the paradigm. Classical Quantum Mechanics (CQM), based on the (nearly heretical) postulate that electrons are physical objects that obey classical physical laws clearly is able to produce, at a minimum, values for ionization energies, excitation energy states for electrons, and even cross sections for scattering, that agree remarkably well with measured values [1-3]. Moreover, these calculations are based on classical physical laws, use very simple algebra, only four physical constants, and no adjustable parameters.

As described in the literature [4], using simple Newtonian force balances, the ionization energy of all known two electron systems (17 values available) is predicted with a single algebraic expression, with one variable; the number of protons. The calculations are based on modeling the electron as a spherical 'bubble', infinitely thin, that surrounds the nucleus. Setting up a Newtonian force balance to describe this physical model is simple as it reduces to a one dimensional problem. Only three measured constants are used in the derived formula: Plank's constant, the permittivity of free space, and the elementary unit of charge on an electron. This is an excellent example, albeit a very elementary one, of the theory because there is no wave equation, only a simple Newtonian force balance is used, the highest level of mathematics required is high school algebra, only the most elementary and thoroughly measured physical constants are employed, there are no fudge factors, and one algebraic equation can be used to solve every known system. There is nothing even marginally as successful in standard quantum theory. Indeed, every effort to solve multi-electron systems is replete with questionable math, and generally serious deviations from true quantum theory [4].

One of the most controversial aspects of CQM is the 'prediction' of the mathematics of the theory that hydrogen smaller than the 'smallest' hydrogen atom can, and should, exist. To wit: there should be states of hydrogen in which the electron radius is smaller than that of the 'ground state' radius (which according to CQM theory is precisely the Bohr radius). Moreover, there is an energy loss associated with this shrinkage of normal hydrogen to one of these 'hydrino' states.

How does one resolve the validity of a theory? This is a complex matter. Physics is not about an invented 'space' such as mathematics, thus, it is not possible to rigorously 'prove' a physics theory. (One cannot prove that gravity will 'work' tomorrow.) It is possible, however, to disprove a theory. The simplest means to disprove a theory is to determine what observable behavior predicted by the theory, and then to test those predictions by looking for the observables. If behavior predicted by the theory is not observed, then the theory is invalidated.

Recently, CQM supporters suggested a theory for the widely observed [5-23] selective broadening of Balmer lines in RF generated Ar/H2 plasmas. The CQM postulate was based on the notion that the line broadening was created by selective transfer of energy to H atoms during hydrino formation. Indeed, Ar atoms, as described in several CQM theory papers, in the RF plasmas are capable of 'catalyzing' the non-spontaneous conversion of H atoms into sub-ground state, shrunken hydrogen atoms, called 'hydrinos'. It was postulated that the energy released by this catalytic process, was transferred directly or indirectly to normal, ground state (as normally defined), H atoms. That is, the large drops in energy resulting from the postulated creation of hydrinos in some cases was converted to kinetic energy of H atoms, rather than photons. Several expected observables are implicit in this application of CQM to the observation of selective Balmer line broadening in Ar/H2 plamsa. Specifically: i) Balmer line broadening should be observed in other 'catalytic' mixtures particularly He/H2 and pure water plasmas, ii) Balmer broadening should not be observed in certain other plasmas (e.g. H2/Xe), iii) Balmer broadening in all plasmas should be independent of angle of observation relative to applied field, iv) Balmer broadening should be observed throughout the plasma, even in low field regions, and v) Balmer broadening should be observed in plasmas generated with other energy sources, including microwave. It is notable that conventional physics models of selective Balmer broadening actually require the acceleration of H atoms to take place in high field gradient regions, and that the H atom velocity should be highest parallel to the field.

A number of experiments, designed to look for these expected observable, have now been reported in the open literature. Below we provide a review of the data collected in some of this work, and report that the line broadening observed is fully consistent with that predicted by the CQM model. Thus it is concluded that line broadening studies have failed to disprove the CQM model. Also critiqued are alternative 'main stream' physics explanations for the many reported observations. The data appear to 'disprove' those models.

## 2. Results and discussion

There are three main sets of results. The first set [24-26] are simply the earliest reports of selective Balmer series line broadening. The discovery of selective Balmer series line broadening appeared to occur 'late' and was unexpected. That is, the rather 'late' reports of selective

Balmer broadening raise a question: Was selective Balmer line broadening observed earlier and simply never reported? This in turn raises another question: Do scientists often fail to report observations that are difficult to explain? The second set of results, Field Affect Models, are reports from laboratories certain that all the fundamentals of physics are well known and that a standard physics explanation for selective Balmer line broadening will eventually be discovered. The failure of this or that theory of the moment, from this perspective, is not a matter of concern. In time, the truth, not requiring any new physics, will be discovered. The third set of experiments were performed by teams attempting to disprove the CQM model. This set is most remarkable of all because of these efforts, surprisingly, failed to disprove this model.

Early Studies: The first reported results were from W. Benesch and E. Lee [24]. They clearly established that dramatic line broadening was found in the Balmer lines of pure hydrogen plasmas generated with a DC glow discharge. They observed hydrogen atoms with energy greater than 100 eV in some cases. Their results also appear to be ambiguous regarding the FA models (more below). Specifically, all the spectral lines were symmetric (shown), but they claim to have detected (not shown) a 'blue shift', consistent with acceleration in the field. Subsequently, no other group has reported a spectral shift. The former is inconsistent with the FA models, and the latter does not appear to figure in any FA theory. In any event, the superposition of a shift due to acceleration of ions by a field, and a broadening due to energy transfer during hydrino formation, is not inconsistent with CQM theory (more below).

Field Affect Model- As noted above, there are two categories of groups that report selective Balmer series line broadening in $H_2$/Ar plasmas. The larger group subscribes to a conventional model of the phenomenon [11-18, 21,22,24,25]. These models include several key features. First, they note only Doppler shifts can explain the remarkably wide spectral lines of hydrogen. Specifically, they interpret the data to indicate H atoms are moving at a velocity orders of magnitude faster than any other atomic or molecular species in the plasma. This can also be expressed as a distinct energy (ca. 30 eV) or temperature (e.g. > 300,000K) for the H atoms. Hence the FA models are based on means to explain how H atoms become hot enough to move with extreme velocities.

It is not the purpose of this paper to review all the ' variations on a theme' of the FA models. That would be a lengthy discussion and the detail required would only distract from the principle points of this essay. The interested reader is referred to the most authoritative papers on the topic [27,28]. However, it is of fundamental importance to the theme of this essay to review features common to all the FA models that implicitly, or explicitly, predict certain behavior. Experiments can be designed on the basis of these 'predicted' behaviors to 'test' the validity of the entire family of FA models.

All FA models include a three step mechanism for hot H atom production: Field acceleration of ionic H species (generally $H_2+$), ii) electron capture due to collision with Argon, and iii) emission. On the basis of this model several observables are expected according to the supporters of FA models. One: There should be a distinct orientation effect [22]. That is, viewed from a position perpendicular to the direction of electric field the average Doppler broadening should be significantly less than that observed along an axis parallel to the field. Two: the expectation is that the effect should only be observed in high field gradient regions. Three: There is no expectation that the broadening should be observed in any but pure $H_2$ or $H_2$/Ar plasmas. Four: the effect should never be observed in a microwave system because ions are two 'heavy' to follow the field, and move sufficiently, to gain energy at microwave frequencies.

Each of the points above can be discussed relative to observation. For example, (Figure 1) in a relatively recent study, Cvetanovic, *et al*. [22] focused on studying line broadening parallel and perpendicular to the field in a capacitively coupled RF system. No evidence was found that persuasively showed a difference. The only data on the topic reported in the study is that found in Figure 1 (Panel C), and unfortunately it was left to the reader to measure the broadening of lines obtained along the field axis. A careful measurement (accounting for the enlargement of the scale of Panel C relative to that of Panel A and B) indicates line broadening perpendicular to the field is virtually the same as that obtained parallel to the field. Regarding the second point above, line broadening in low field gradient regions: Workers performing studies in support of the FA model do not appear to have collected any data away from high field gradient regions. A review of the work conducted by supporters of the FA model suggests these workers (inadvertently) made an assumption: *There is no reason to search for line broadening away from the high field region. Line broadening cannot exist away from regions of high field gradient.* This assumption proved to be wrong.

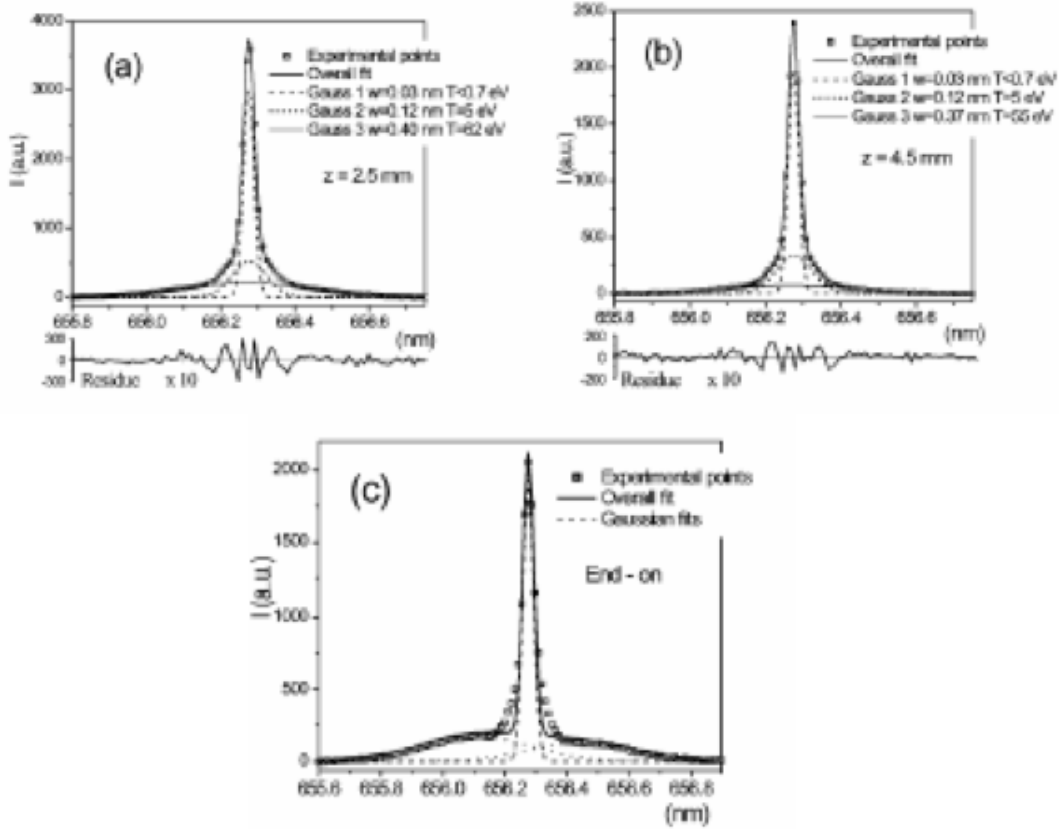

FIGURE 1: Hα Line Broadening in RF Plasma. This figure, taken from a recent publication (Ref. 22) of supporters of the FA model family, shows the magnitude of broadening is not a function of angle of observation relative to the electric field direction in the high field gradient region of a low pressure (<1 Torr) RF plasma.

This group of researchers also never reports studies of line broadening in gas mixtures other than hydrogen and argon. In some versions of the FA model Ar is a key ingredient. Ar is postulated to have a particularly high cross section for charge exchange with $H_2$ ions, thus argon is required for that step in all FA models in which the accelerated hydrogen ion captures an electron (see above). In other words, there are no reports of control studies required to experimentally assess the role of argon.

This group of workers also performed limited studies of line broadening in microwave generated plasmas, and was unable to observe any. However, it is clear they searched a very limited region of parameter space that in that work [29], and that the parameter space 'searched' did not overlap that in which others observed line broadening in microwave plasmas [30,31]. Workers reporting line broadening in microwave plasmas report many 'failures' as well, suggesting that precise parameters, and particular equipment (e.g. Evanson coupler) is required.

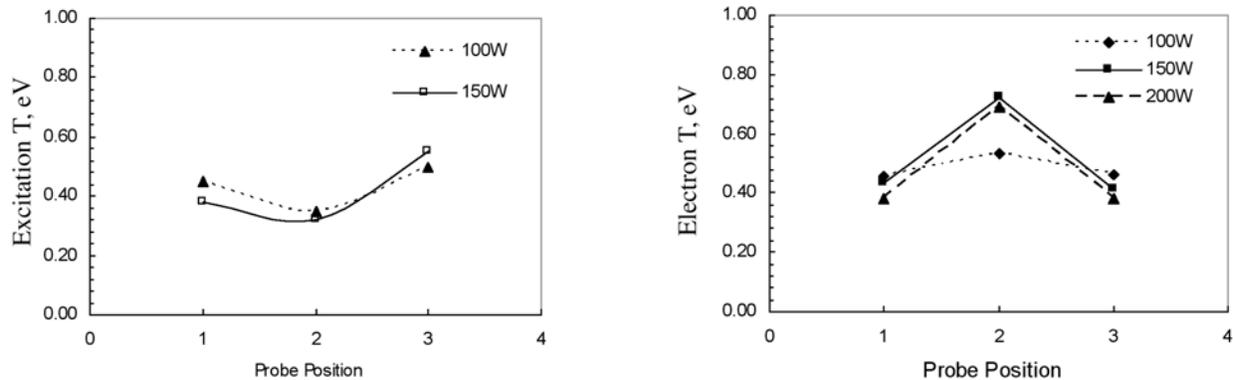

FIGURE 2: Temperature: Excitation and Electron. Detailed measurements of electron and excitation temperatures in H2/Ar in low pressure (<1 torr) RF generated plasmas never show energies even as high as 1 eV (From Ref. 5). Note, position 2 is in the high field region between capacitive coupling plates, position 1 is at the gas entry end 15 cm from plates and position 3 at the gas exit end, also 15 cm from the high field.

In sum, supporters of the FA model performed little work designed to test the underlying principles of their models. In those cases in which they did conduct studies, the results did not support the FA family of models.

CQM Tests- The third group of workers are those attempting to help validate the CQM model of fast hydrogen production, via catalytic production of hydrinos, by the standard scientific method: Attempting to disprove it [32]. That is, they designed experiments to debunk the CQM fast hydrogen/hydrino formation models. In particular, they focused their attention on looking for features of plasmas predicted to exist by the CQM model, but predicted not to exist according to the FA models. Failure to observe these features would certainly be a blow to the CQM model. Observing these features would allow CQM to 'survive' as a valid model, but would not/could not constitute proof.

In particular, these workers have looked for these five features of plasmas predicted by CQM. One: There is no orientation effect. As the CQM model of fast H production is based on a 'chemical process' the energy of fast H atoms should not be a function of the angle of orientation relative to the applied field. Two: Fast H atoms should be found throughout the plasma. Again, acceleration of ionic species by a field is not part of the CQM model. Hence, the process can take place regardless of the field strength. All that is required is relatively high concentrations of catalyst (e.g. He+) and H atoms. High energy H atoms should be found throughout the volume of plasma, both near a region of high field gradient and in areas in which there is virtually no field gradient. Three: Selective line broadening should not only occur in pure $H_2$ and Ar/$H_2$ plasmas, but also in pure water plasmas and He/$H_2$ plasmas. As a corollary, it should not be observed in certain plasmas such as Xe/$H_2$ plasmas. Four: As the process is chemical in nature, and does not require ion acceleration, broadening should be found in microwave plasmas as well. Five: Some specific, high energy, spectral lines should be found in the EUV spectra of mixed gas plasmas (e.g. He/$H_2$) that are never found in either gas independently.

There is a fairly large body of literature that bears on the topic of observations of plasma features consistent with the predictions of CQM, albeit most come from the laboratory of the author of this report [5], from the laboratory of Black Light Power [6-10,23,30,34], or from a collaboration between the two [5,31,35]. This paper focuses, but not exclusively, on those studies involving the author. It is clear that reports from other laboratories will be reported in the near future.

The most thorough study of the first predicted effect (no correlation to field direction) was our study of broadening as a function of orientation relative to the electric field in capacitively coupled RF plasma (Figure 3). It was clearly shown that there was virtually no orientation effect. In fact, if there was a slightly larger broadening observed, it was observed along a line perpendicular to the applied field, completely contradicting the 'expectation' of the FA family of models.

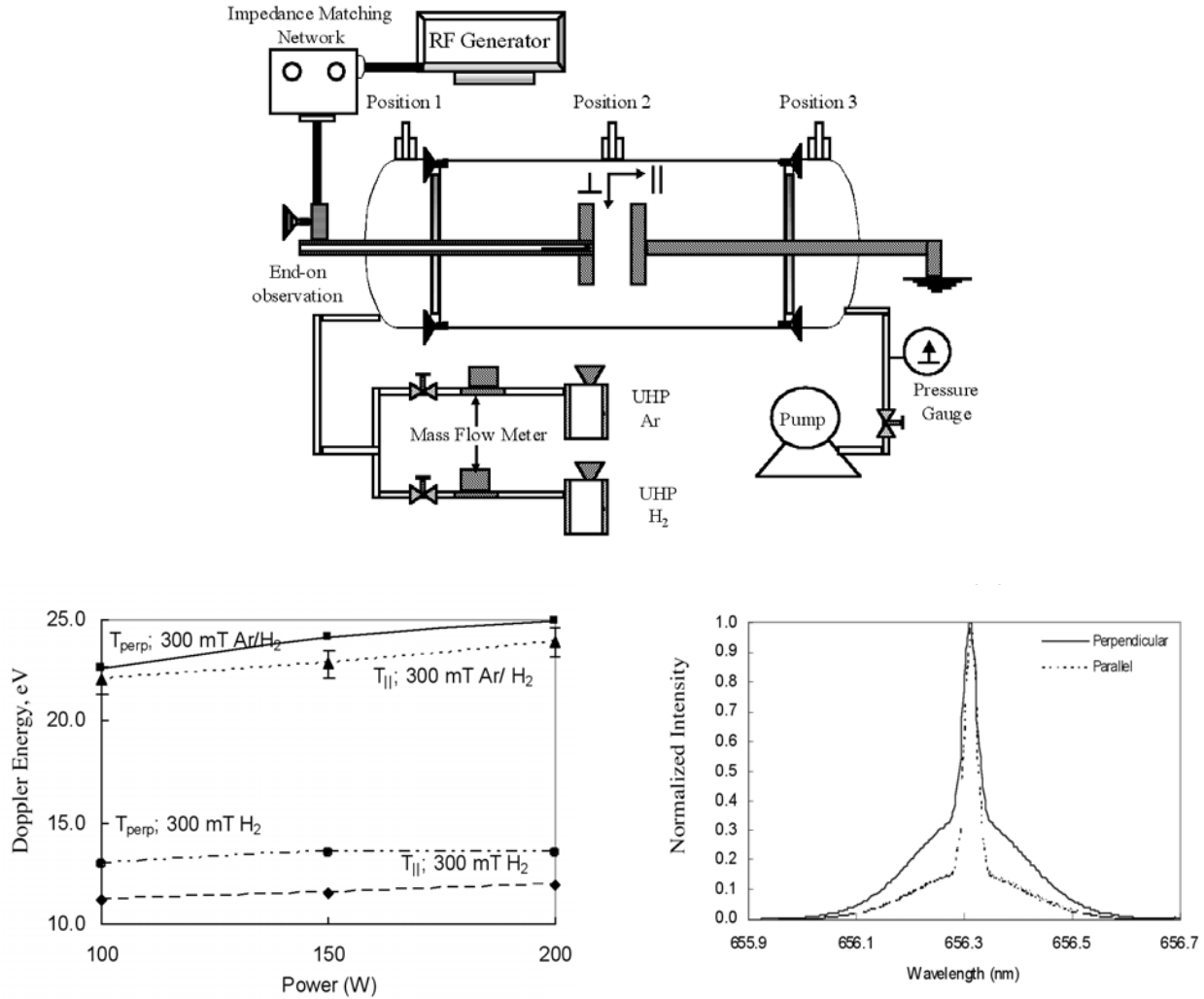

FIGURE 3- Parallel vs. Perpendicular. The TOP panel shows the apparatus used. The RF cavity is approximately 15 cm in diameter and 35 cm. long, with two plates for capacitive coupling of RF (high field region) about one centimeter apart in the center. It is set up so that spectra can be taken parallel to the field by inserting a light fiber from the left side to a quartz window in the center of the plate. The BOTTOM LEFT panel shows there is virtually no orientation effect in either $Ar/H_2$ or pure $H_2$ plasmas. The BOTTOM RIGHT panel shows typical $H_\alpha$ lines. (From Ref. 5)

We also found that line broadening was found precisely in those RF capacitively coupled plasmas predicted by the CQM model: $Ar/H_2$ (5), $He/H_2$ (Ref. 33 ,Figure 4) and pure $H_2O$ plasmas (Refs. 34 and 35, Figure 5). Also, no line broadening was found in the same RF chambers, with identical operating parameters used for $Ar/H_2$, $He/H_2$ and $H_2O$ plasmas, except $Xe/H_2$ gas mixes were employed.

Studies from our lab (Figure 4, Refs. 5,33,35) also make it clear that line broadening is found throughout the plasma, and that it is equal in magnitude to the line broadening in the high field region. Yet, it is quite clear that there is a low field outside the discharge region, and that away from the discharge there is virtually no field gradient.

Our team also produced a significant volume of data showing that selective line broadening is also found in microwave plasmas [31], particularly with water plasmas (Figure 6). Other laboratories, using different microwave couplers, and gas 'mixtures' other than water, report only moderate, but measurable, selective Balmer line broadening in microwave generated plasmas [36-38]. It is reasonable to anticipate that selective Balmer line broadening in microwave plasmas will be an area of intense effort by many teams in the near future.

Finally, Mills team has published a number of papers in which EUV spectral lines predicted by CQM are found in mixed gas plasmas. Careful control studies with only one gas show no such lines [39,40]. There are also extensive published calorimetric [41] and NMR studies [42] consistent with CQM.

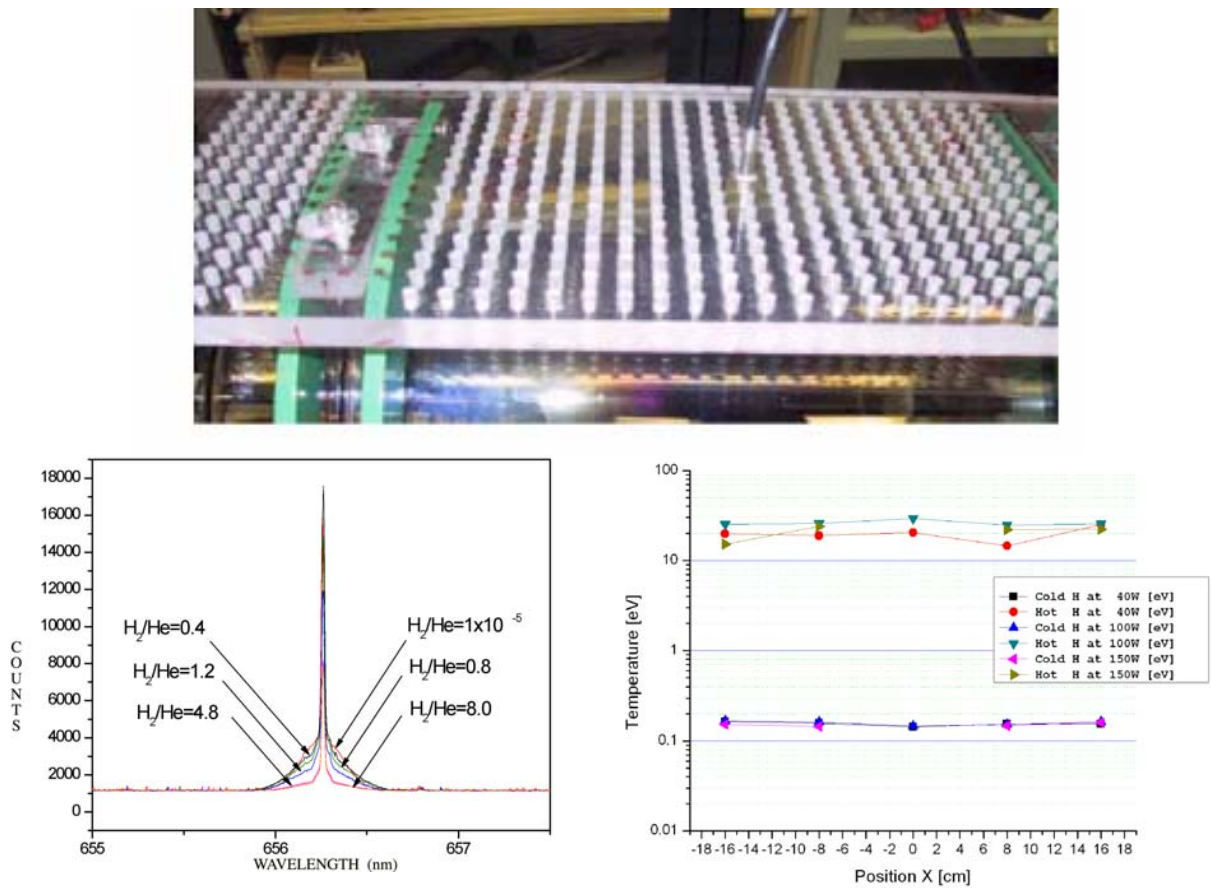

FIGURE 4- RF He/$H_2$ Plasma. TOP- Data were primarily collected from above, perpendicular to the electric field. BOTTOM LEFT- Typical data. The fraction of H in the hot state increased with a decrease in H2/He ratio. BOTTOM RIGHT- The magnitude of broadening was little impacted by position, or applied power. (From Ref. 33)

Is the CQM model consistent with thermodynamics? Yes. The H atoms are not heated via a thermal transfer mechanism. That is, the 'engine' of H atom heating is not a thermal one. Rather, there is a direct chemical reaction process that heats the H atoms directly.

In sum, the RF plasma experiments of the CQM proponents were successful, in that a simple summary is possible: *All outcomes were consistent with the predictions of CQM and inconsistent with all FA models.* This does not imply that this work proves the CQM model. However, at a 'lower bound' these workers, including our lab, failed to disprove the CQM model, and have shown that no earlier line broadening data is inconsistent with it.

## 3. Conclusion

A thorough review of studies of selective Balmer series line broadening shows none of the data is inconsistent with the novel CQM model. In contrast, a great deal of the data appears to be 'unanticipated' on the basis of conventional physics models. In particular, consistent with CQM and inconsistent with conventional attempts to explain selective Balmer line broadening, the following observations have been made: First, the extent of line broadening is independent of angle of observation relative to the electric field gradient. Second, line broadening, of a consistent magnitude, is found throughout plasmas, not just in high field gradient regions. Third, line broadening is found in all those mixtures predicted to generate line broadening by the CQM model. Fourth: selective Balmer line broadening has also been found in microwave plasmas.

In addition to providing experimental data to distinguish between two theories, the He/H2 the data is arguably of intrinsic significance. Indeed, He/$H_2$ plasmas are among the most common structures of matter in the universe. If an interaction between these two gases in a pressure (a) and 1.0 Torr (b). Very little line broadening was found when Evenson coupler replaced with an RF coil and system operated with water plasma 0.2 Torr (c). No line broadening was found for the oxygen radical (3971 Å) line for either the Evenson (d) or RF (e) water plasmas operated at 0.2 Torr. (From Ref. 31) plasma leads to the selective generation of extremely energetic hydrogen (not helium), even in the absence of a field, then this phenomenon needs to be thoroughly investigated. In this regard, some interesting and fascinating data already exists. To wit: Balmer series line broadening is also observed in many stellar spectra, including flares [43,44] and some particular star classes (e.g. cool Ap stars, A and F dwarfs [45-47]). In flares, as in most laboratory systems, it is clear that there are two types of hydrogen. That is the spectral lines are composed of two components, a central line, which is only modestly broadened, and 'wings' which are dramatically broadened. Most astrophysics interpretations assume that the modest broadening of the central line comes from Doppler broadening, and the wings come from Stark broadening.

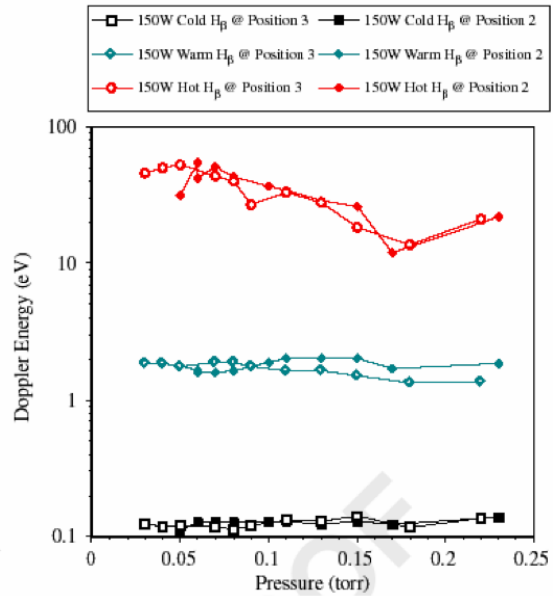

FIGURE 5 - Line Broadening in Pure H2O. H$\alpha$ lines collected at 150 W between the plates were fit with three Gaussians. It is clear that there are some extremely hot H atoms present in these low pressure RF generated plasmas (From Ref. 35).

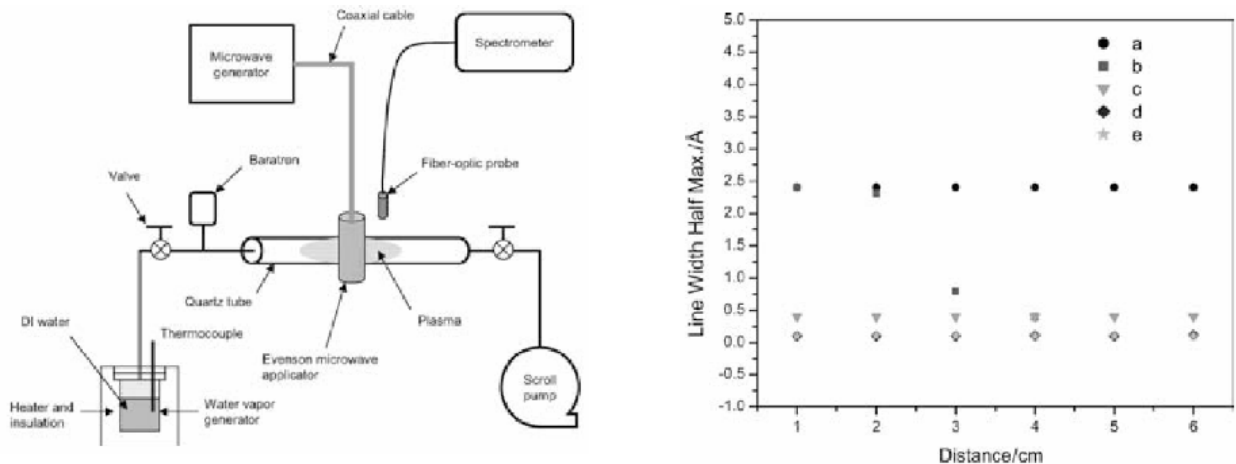

FIGURE 6. Line Broadening at 2.45 GHz. LEFT- Line broadening was found in restricted cases: Pure water, Evenson coupler, ca. 1 Torr pressure, ~1 cm diameter quartz tube. RIGHT-Evenson cavity, 0.2 Torr

The charge concentrations required to produce the observed broadening are enormous, but perhaps these concentrations can be achieved in stars. (For example, assuming a fraction ionized of 1 in 10,000 typical of plasmas at 5000 K, and the ideal gas law, pressures of 50,000 atmospheres would be required to create the needed charge densities. And the overall density would be 'solid'-like, at 1 atom/$10Å^3$.) However, to explain the observed excessive broadening in terrestrial plasmas on the basis of the Stark effect would require charge concentrations in these plasmas $10^4$ times higher than the atomic concentration. Moreover, the high charge concentration would broaden all spectral lines, not selectively broaden only hydrogen spectral lines, as discussed in detail in this review. Clearly Stark broadening cannot explain the broadening observed in terrestrial plasmas. It is puzzling to note that the reports of spectral line broadening in astrophysics are specifically tied to hydrogen spectral lines, and other lines, when analyzed, are not found to show any indication of Stark broadening [43].